\documentstyle [11pt]{article}

\makeatletter

\newcommand{\singlespacing}{\let\CS=\@currsize\renewcommand{\baselinestretch}{1}\tiny\CS}
\newcommand{\oneandahalfspacing}{\let\CS=\@currsize\renewcommand{\baselinestretch}{1.25}\tiny\CS}
\newcommand{\doublespacing}{\let\CS=\@currsize\renewcommand{\baselinestretch}{1.35}\tiny\CS}

\def\@citex[#1]#2{\if@filesw\immediate\write\@auxout{\string\citation{#2}}\fi
  \def\@citea{}\@cite{\@for\@citeb:=#2\do
    {\@citea\def\@citea{,\linebreak[0]\hskip0pt plus .2em}%
      \@ifundefined{b@\@citeb}%
      {{\bf ?}\@warning{Citation `\@citeb' on page \thepage\space undefined}}%
      \hbox{\csname b@\@citeb\endcsname}}}{#1}}

\newtheorem{rule-def}[theorem]{Rule}

\begin{document}
\newcommand{\la}{\lambda}
\newcommand{\si}{\sigma}
\newcommand{\ol}{1-\lambda}
\newcommand{\be}{\begin{equation}}
\newcommand{\ee}{\end{equation}}
\newcommand{\bea}{\begin{eqnarray}}
\newcommand{\eea}{\end{eqnarray}}
\newcommand{\nn}{\nonumber}
\newcommand{\lb}{\label}

\begin{center}
{\large \bf THE DYNAMIC COSMOLOGICAL TERM $\Lambda$: SOME ASPECTS OF
PHENOMENOLOGICAL MODELS}
\end{center}

\begin{center}
  UTPAL MUKHOPADHYAY$^1$ and SAIBAL
RAY$^2$\\
$^1${\it Satyabharati Vidyapith, North 24 Parganas, Kolkata
700 126, West Bengal, India\\$^2$Physics Department, Barasat
Government College, Kolkata 700 124, North 24 Parganas, West
Bengal, India \& Inter-University Centre for Astronomy and
Astrophysics, PO Box 4, Pune 411 007, India;
E-mail:saibal@iucaa.ernet.in}
\end{center}
{\bf Abstract}. Choosing a phenomenological model of $\Lambda$,
viz. $\Lambda \sim \dot H$, it has been shown that this model of
$\Lambda$ is equivalent to other three types of $\Lambda$,
$\Lambda \sim (\dot a/a)^2$, $\Lambda \sim \ddot a/a$ and $\Lambda
\sim \rho$. Through an indirect approach, it has also been
possible to put a limit on the deceleration parameter  $q$. It has
been shown that if $q$ becomes less than $-1$, then this model can
predict about the presence of phantom energy. \\

PACS number(s): 98.80.Jk, 98.80.Cq.\\

\begin{center}
{\bf 1. Introduction}
\end{center}
 The experiments on supernovae type Ia have been
provided conclusive evidences that the present expanding Universe
is in accelerating phase (Perlmutter et al. 1998; Riess et al.
1998). Obviously, this suggests that a repulsive gravitation, in
the form of some kind of exotic energy, is acting behind this
unexpected phenomenon. The so-called cosmological constant
$\Lambda$ was thought to be a possible candidate of this exotic or
dark energy. However, $\Lambda$ was assumed as a dynamic term
depending on the cosmic time so that it can explain the rate of
acceleration.

As a consequence of the search for current status of this
acceleration some phenomenological models of $\Lambda$, viz.
$\Lambda \sim (\dot a/a)^2$, $\Lambda \sim \ddot a/a$ and $\Lambda
\sim \rho$, recently we (Ray and Mukhopadhyay 2004) have shown the
equivalence of these models. We also have established a
relationship between the parameters $\alpha$, $\beta$ and $\gamma$
of the respective models. It was mentioned that $\Lambda \sim
\ddot a/a$ model can be viewed as a combination of $\Lambda \sim
(\dot a/a)^2$ and $\Lambda \sim \dot H$ models (since, $\ddot a/a
= (\dot a/a)^2 + \dot H$). Therefore, for $\dot H = 0$ the models
$\Lambda \sim H^2$ and $\Lambda \sim \ddot a/a$ become identical,
where $a(t)$ is the scale factor of the Universe and $H(=\dot
a/a)$ is the Hubble parameter. Since, $\Lambda \sim \ddot a/a$
model depends on $ H^2$ and $\dot H$, and $\Lambda \sim H^2$ model
has already been studied by us (Ray and Mukhopadhyay 2004), the
point of interest is now $\Lambda \sim \dot H$ model. Although a
number of phenomenological models have been listed by Overduin and
Cooperstock (1998) (also see the references in Ray and
Mukhopadhyay 2004) but $\Lambda \sim \dot H$ model is not included
there.

In the present paper, Einstein's field equations have been solved
for $a(t)$, $\rho(t)$ and $\Lambda(t)$ with the {\it ansatz}
$\Lambda= \mu \dot H$, where $\mu$ is a parameter of $\Lambda \sim
\dot H$ model. It has been shown that $\Lambda \sim \dot H$ model
is equivalent to $\Lambda \sim (\dot a/a)^2$, $\Lambda \sim \ddot
a/a$ and $\Lambda \sim \rho$ models when the solutions for a
$a(t)$, $\rho(t)$ and $\Lambda(t)$ are expressed in terms of
$\Omega_{m}$ and $\Omega_{\Lambda}$, respectively the matter- and
vacuum-energy densities of the Universe. It has also been possible
to connect $\mu$ with $\alpha$, $\beta$ and $\gamma$. Section 5
deals with the physical implications and limit on the deceleration
parameter $q$, for the present model while equivalence of $\Lambda
\sim \dot H$ model with $\Lambda \sim (\dot a/a)^2$, $\Lambda \sim
\ddot a/a$ and $\Lambda \sim \rho$ models as well as
interrelationship of $\mu$ with $\alpha$, $\beta$ and $\gamma$ are
presented in Section 4. Sections 2 and 3 deals, respectively, with
Einstein's field equations and their solutions. Finally, all the
results and their significance are discussed in Section 6.\\

\begin{center}
{\bf 2. Einstein Field Equations}
\end{center}
\noindent
The Einstein field equations are given by
 \bea
  R^{ij} - \frac{1}{2}Rg^{ij} = -8\pi
G\left[T^{ij} - \frac{\Lambda}{8\pi G}g^{ij}\right] \eea where
$\Lambda = \Lambda(t)$ is the time dependent cosmological constant
and $c$, the velocity of light in vacuum, is assumed to be unity
in relativistic units. If we choose the spherically symmetric
Friedmann-Lema{\^i}tre-Robertson-Walker (FLRW) metric \bea ds^2 =
-dt^2 + a(t)^2\left[\frac{dr^2}{1 - kr^2} + r^2 (d\theta^2 +
sin^2\theta d\phi^2)\right] \eea where the curvature constant $k$
can assume the values $ -1, 0, +1 $ respectively for open, flat
and close Universe models, then for a flat Universe, equation (1)
yields respectively the Friedmann equation and the Raychaudhuri
equation which are given by \bea
 3\left(\frac{\dot a}{a}\right)^2
= 8\pi G \rho + \Lambda,
\eea
\bea
3\frac{\ddot a}{a} = - 4\pi
G(\rho + 3p)+ \Lambda.
\eea
Let us now consider the barotropic
equation of state in the form
 \bea
  p = w \rho
\eea
 where  $w$, the equation of state parameter, for the dust, radiation, vacuum
 fluid and stiff fluid can take the constant values $0$, $1/3$, $-1$ and $+1$
respectively.

Let us assume that $G$ does not vary with space and time. Then for
the ansatz \bea \Lambda= \mu \dot H \eea where $\mu$ is a free
parameter, we get from the equations (3) and (4) the following two
modified equations
 \bea
 4\pi G\rho =
\frac{1}{2}(3H^2-\mu \dot H).
 \eea
 \bea
3(H^2 + \dot H)= -4\pi G\rho (1+3\omega)+ \mu \dot H
 \eea
Equations (7) and (8), on simplification, yield the differential
equation
 \bea
 (2-\mu-\mu\omega)\dot H=-3(1 + w)H^2.
 \eea

\begin{center}
{\bf 3. The Models}
\end{center}
\noindent 
  Now, equation (9) on integration gives \bea
 H= \frac{2-\mu-\mu\omega}{3(1 + w)t}.
 \eea
 Putting $H= \dot a/a$ in equation (10) and integrating it further
 we get our general solution set as
  \bea
  a(t) = Ct^{\frac{2-\mu-\mu\omega}{3(1 + w)}},
  \eea
  \bea
  \rho(t) = \frac{2-\mu-\mu\omega}{12\pi G(1 + w)^2}t^{-2},
  \eea
   \bea
\Lambda(t) = \frac{-\mu(2-\mu-\mu\omega)}{3(1 + w)}t^{-2} \eea
 where $C$ is an integration constant.\\

\noindent
 (i) {\it Dust case ($\omega=0$)}\\
  For dust case, equations (11), (12), (13) and (10), respectively takes the form
\bea a(t) = Ct^{(2-\mu)/{3}},
 \eea
 \bea
 \rho(t) = \frac{2-\mu}{12\pi G}t^{-2},
\eea
 \bea
 \Lambda(t) = \frac{-\mu(2-\mu)}{3}t^{-2},
 \eea
\bea
 t = \frac{2-\mu}{3H}.
\eea Equation (15) suggests that for physically valid $\rho$ (i.e.
$\rho>0$), we should get $\mu<2$. So, $\mu$ can be negative as
well. Again, from equation (16) we find that for a repulsive
$\Lambda$, the constraint on $\mu$ is that it must be negative.
Thus, equations (15), (16) and (17) all point towards a negative
$\mu$. \\
\noindent
 (ii) {\it Radiation case ($\omega=1/3$)}\\
  For radiation case, equations (11), (12), (13) and (10), respectively reduce to
\bea a(t) = Ct^{(3-2\mu)/6},
 \eea
 \bea
 \rho(t) = \frac{3-2\mu}{32\pi G}t^{-2},
\eea
 \bea
 \Lambda(t) = \frac{-\mu(3-2\mu)}{6}t^{-2},
 \eea
\bea
 t = \frac{3-2\mu}{6H}.
 \eea
Equation (19) suggests that for physically valid $\rho$, $\mu<3/2$
whereas equation (20) demands a negative $\mu$ for repulsive
$\Lambda$. Thus, in this case also a negative $\mu$ is necessary
for equations (19) - (21).\\

\begin{center}
{\bf 4. Equivalent Relationship Between $\Lambda$-Dependent Models}
\end{center}
\noindent
  Using the value of $t$ from equation
(17) in equation (15) we get \bea
 \Omega_{m} = \frac{2}{2- \mu},
\eea where $ \Omega_{m}(=8\pi G\rho/3H^2)$ is the matter-energy
density of the Universe.

Again, using equation (17) we get from equation (16)
\bea
 \Omega_{\Lambda} = \frac{-\mu}{2- \mu},
\eea
where $ \Omega_{\Lambda}(=\Lambda/3H^2)$ is the vacuum-energy
density of the Universe.

Adding (22) and (23) we get,
\bea
 \Omega_{m} + \Omega_{\Lambda} = 1
 \eea
 which is another form of equation (3) for flat Universe. Also,
 using the value of $\mu$ from equation (22), we get from (17),
\bea
 t = \frac{2}{3\Omega_{m}H}.
 \eea
Thus, if $t_0$ and $H_0$ be the values of  $t$ and $H$ at the
present epoch, then
\bea
 t_0 = \frac{2}{3\Omega_{m0}H_0}.
 \eea
Equation (26) is the same expression for $t_0$ as obtained by us
(Ray and Mukhopadhyay 2004) for $\Lambda \sim (\dot a/a)^2$,
$\Lambda \sim \ddot a/a$ and $\Lambda \sim \rho$. Again, using
equation (23) in equations (14)-(16), it can easily be shown that
the expressions for $a(t)$, $\rho(t)$ $\Lambda(t)$ become
identical with the expressions for those quantities as obtained by
us (Ray and Mukhopadhyay 2004). This means that $\Lambda \sim \dot
H$ model is equivalent to $\Lambda \sim (\dot a/a)^2$, $\Lambda
\sim \ddot a/a$ and $\Lambda \sim \rho$. \noindent Again, from
equation (22) we get, \bea
 \mu= - \frac{2\Omega_{\Lambda}}{\Omega_{m}}.
 \eea
But, we (Ray and Mukhopadhyay 2004) already have shown that \bea
 \gamma= \frac{\Omega_{\Lambda}}{\Omega_{m}}.
 \eea
Thus, equations (27) and (28) yield \bea
 \mu= - 2\gamma.
 \eea
Also, we (Ray and Mukhopadhyay 2004) have shown that for
pressureless dust $\alpha$, $\beta$, $\gamma$ respectively the
three parameters of $\Lambda \sim (\dot a/a)^2$, $\Lambda \sim
\ddot a/a$ and $\Lambda \sim \rho$ models can be interconnected by
the relation \bea \alpha = \frac{\beta}{3(\beta - 2)}=
\frac{\gamma}{\gamma+1}. \eea Thus, combining equation (29) with
equation (30) we have, \bea \alpha = \frac{\beta}{3(\beta - 2)}=
\frac{\gamma}{\gamma +1}= \frac{\mu}{\mu-2}.
 \eea
Similarly, it can be shown that for radiation-filled Universe,
$\mu$ is related to $\Omega_{m}$ and $\Omega_{\Lambda}$
respectively by the relations,
\bea
\Omega_{m}=\frac{3}{3-2\mu},
 \eea
\bea
\Omega_{\Lambda}=\frac{-2\mu}{3-2\mu}.
 \eea
 Adding (32) and (33) we get again,
\bea
 \Omega_{m} + \Omega_{\Lambda} = 1.
 \eea
In this case, $\mu$ is related to $\gamma$ of our previous
investigation (Ray and Mukhopadhyay 2004) by \bea
 \mu= - \frac{3}{2}\gamma.
 \eea
Using the value of $\mu$ from equation (32) in equation (21) we
have, \bea
 t = \frac{1}{2\Omega_{m}H}
 \eea
which for the present Universe yields
\bea
 t_0 = \frac{1}{2\Omega_{m0}H_0}.
 \eea
Again, equation (37) is the same expression for $t_0$ as obtained
by us (Ray and Mukhopadhyay 2004) for $\Lambda \sim (\dot a/a)^2$,
$\Lambda \sim \ddot a/a$ and $\Lambda \sim \rho$ models in
radiation case. Also, combining equation (35) with equation (39)
of our previous investigation (Ray and Mukhopadhyay 2004) we find
that in the radiation case, $\mu$ is related to $\alpha$, $\beta$,
$\gamma$ by
 \bea
 \alpha = \frac{\beta}{2\beta - 3}= \frac{\gamma}{\gamma +1}=
\frac{2\mu}{2\mu-3}.
 \eea

\begin{center}
{\bf 5. Physical Features of the Models}
\end{center}
\noindent 
 The
deceleration parameter $q$ is given by
 \bea
 q = - \frac{a\ddot a}{\dot a^2}= -(1+ \frac{\dot H}{H^2}).
\eea Thus, using equation (10) we have \bea q=
-\left[1-\frac{3(1+\omega)}{2-\mu-\mu\omega}\right]. \eea For an
accelerating Universe, $q<0$ and hence,
$\mu<-(1+3\omega)/(1+\omega)$. Equation (40) tells us that for a
dust-filled accelerating Universe, $\mu$ should be less than $-1$.
We have already shown that for physically valid $\rho$, $\Lambda$
and $t$, $\mu$ must be negative. Thus, for $\mu<-1$ we get an
accelerating Universe with repulsive $\Lambda$ through our
$\Lambda \sim \dot H$ model.

Again, equation (39) can be written as
\bea
\dot H= -H^2(q+1)
 \eea
so that one can find an expression for the variable cosmological
constant as
 \bea
 \Lambda = -\mu H^2(q+1)
 \eea
 as $\Lambda= \mu \dot H$. If $\mu<0$, then equation (42) tells us
 that $\Lambda$ remains a repulsive force (i.e. $\Lambda>0$ so long as
 $q>-1$. This means that if the acceleration of the Universe
 exceeds a certain limit then $\Lambda$ will become an attractive
 force. In the present accelerating Universe, $q$ lies near $-0.5$
 (Overduin and Cooperstock 1998; Efstathiou 1998; Sahni 1999)
  and hence $(q+1)>0$. Since, for our model both
 $\dot H$ and $H^2$ are proportional to $t^{-2}$ then it is clear
 from equation (39) that $q$ is a constant quantity. But through our
 model it has been possible to exhibit indirectly via equation (42)
 that in future $\Lambda$ may become an attractive force provided
 $q$ becomes less $-1$.\\

\begin{center}
{\bf Discussions}
\end{center}
 \noindent
It has been shown, in the present investigation, that $\Lambda
=\mu \dot H$ model is equivalent to other three types of
phenomenological models, viz. $\Lambda =\alpha (\dot a/a)^2$,
$\Lambda =\beta {\ddot a/a}$ and $\Lambda =\gamma \rho$. Also, a
relationship is established between the parameters $\mu$ and
$\gamma$ in both dust and radiation cases. It was observed in the
previous investigation (Ray and Mukhopadhyay 2004) that $\alpha$
and $\gamma$ are related to  $\Omega_{m}$ and $\Omega_{\Lambda}$
through the same relation in both dust and radiation cases while
$\beta$ is related to $\Omega_{m}$ and $\Omega_{\Lambda}$ through
different relations as expression for $\beta$ in terms of these
cosmic energy densities contains $\omega$. Since, ${\ddot a}/{a} =
H^2 + \dot H$, then it is clear that dependency of the parameter
$\omega$ on $\beta$ is due to $\Lambda \sim \dot H$ part because
relation of $\mu$ with $\Omega_{m}$ and $\Omega_{\Lambda}$ also
contains $\omega$ and hence $\mu$ behaves differently with cosmic
matter and vacuum energy density parameters in dust and radiation
cases.

It has been possible through our model to put a limit on $q$, the
deceleration parameter of the Universe. If one is content with the
idea of a repulsive $\Lambda$ only, then we find that $q$ cannot
exceed a certain limit. But, if one is bold enough to think of an
attractive $\Lambda$ in future, then $q$ can go on decreasing
indefinitely, i.e. the rate of acceleration can increase
indefinitely. But in presence of an attractive $\Lambda$,
indefinite increase in acceleration is possible only if $G$ goes
on decreasing with time. Now, an indefinite increase in
acceleration may lead to `Big Rip' (Caldwell et al. 2003) or
`Partial Rip' ({\u{S}}tefan{\u{c}}i{\'c}, 2004a) of the Universe
for a super-negative ($< -1$) equation of state parameter known as
phantom energy. Therefore, although we have worked out the present
model with the assumption that $G$ is constant, yet through an
indirect approach it has been possible to predict about the
presence of phantom energy. Moreover, recently
{\u{S}}tefan{\u{c}}i{\'c} (2004b) has shown that cosmological
model without phantom energy may exhibit the same effect as that
of phantom energy. This is very significant because in the present
work also without any {\it a priori} assumption regarding phantom
energy, it is shown that in future the Universe may go on
expanding with an ever increasing acceleration similar to that of
phantom energy models if the value of the deceleration parameter
drops below $-1$.

Finally, it should be mentioned that any linear combination of
$\Lambda \sim (\dot a/a)^2$, $\Lambda \sim \ddot a/a$ and $\Lambda
\sim \rho$ is also equivalent.\\

\begin{center}
{\bf Acknowledgement}
\end{center}
\noindent
  One of the authors (SR) would like to
express his gratitude to the authorities of IUCAA, Pune for
providing him the Associateship Programme under which a part of
this work was carried out.\\

\begin{center}
{\bf References}
\end{center}
 \noindent
 Caldwell, R. R. et al.: 2003, {\it Phys. Rev. Lett.} {\bf 9}, 071301-1.\\
 \noindent
Efstathiou, G. et al.: 1998, {\it astro-ph/9812226}.\\
 \noindent
Overduin, J. M. and Cooperstock, F. I.: 1998, {\it Phys. Rev. D} {\bf 58},  043506.\\
\noindent
Perlmutter, S. et al.: 1998, {\it Nat.} {\bf 391}, 51.\\
\noindent
Ray, S. and Mukhopadhyay, U.: 2004, {\it astro-ph/0407295}.\\
\noindent
 Riess, A. G. et al.: 1998, {\it Astron. J.} {\bf 116}, 1009.\\
\noindent
Sahni, V.: 1999, {\it Pramana} {\bf 53}, 937.\\
\noindent
 {\u{S}}tefan{\u{c}}i{\'c}, H.: 2004a, {\it  Phys. Lett. B} {\bf 595}, 9.\\ 
\noindent 
{\u{S}}tefan{\u{c}}i{\'c}, H.: 2004b, {\it  Eur. Phys. J. C} {\bf 36}, 523.

\end{document}